\documentclass[letters]{IEEEtran}

\usepackage{color,graphicx,amsmath,amssymb,amsthm,epsfig,mathrsfs,cite,bm,graphics}
\usepackage{dblfloatfix}    
\usepackage{algpseudocode}
\usepackage{algorithmicx}
\usepackage{algorithm} 
\usepackage[utf8]{inputenc}
\usepackage{amsmath}
\usepackage{amsfonts}
\usepackage{amssymb}
\usepackage{placeins}
\usepackage{graphicx}
\usepackage{bbm,bm}
\usepackage{graphicx,graphics,subcaption}
\usepackage{tikz}
\usepackage{multirow}
\usepackage{authblk}
\usepackage{tikz}
\usetikzlibrary{automata, positioning, arrows,shapes.multipart,fit}
\usepackage{pgfplots}
\usepackage[normalem]{ulem}
\usepackage{cancel}
\usepackage{soul}

\usepackage{subcaption}
\usepackage[]{todonotes}
\presetkeys{todonotes}{color=blue!20}{}

\newcommand{\revAdd}[1]{#1}
\newcommand{\revDel}[1]{}

\newcommand{\mymat}[1]{\textbf{#1}}
\newcommand{\myvec}[1]{\textbf{#1}}

\newcommand\blfootnote[1]{%
  \begingroup
  \renewcommand\thefootnote{}\footnote{#1}%
  \addtocounter{footnote}{-1}%
  \endgroup
}

\begin{document}
    \title{Beyond 5G: Leveraging Cell Free TDD Massive MIMO using Cascaded Deep learning}
    \author{Navaneet Athreya, Vishnu Raj and Sheetal Kalyani\vspace{-4ex}}

	\maketitle
	\begin{abstract}
		This paper deals with the calibration of Time Division Duplexing (TDD) reciprocity 
in an Orthogonal Frequency Division Multiplexing (OFDM) based Cell Free Massive MIMO 
system where the responses of the (Radio Frequency) RF chains render the end to end 
channel non-reciprocal, even though the physical wireless channel is reciprocal. 
We further address the non-availability of the uplink channel estimates at locations 
other than pilot subcarriers and propose a single-shot solution to estimate the downlink 
channel at all subcarriers from the uplink channel at selected pilot subcarriers. 
We propose a cascade of two Deep Neural Networks (DNN) to achieve the objective. 
The proposed method is easily scalable and removes the need for relative reciprocity 
calibration based on the cooperation of antennas, which usually introduces dependency 
in Cell Free Massive MIMO systems.
	\end{abstract}
	\blfootnote{Navaneet Athreya is with Department of ECE, Virginia Tech, Blacksburg, VA, USA (e-mail: navaneet@vt.edu). \\
    \indent Vishnu Raj and Sheetal Kalyani are with the Department of Electrical Engineering at IIT Madras, Chennai, India (e-mail: \{ee14d213, skalyani\}@ee.iitm.ac.in).}
	\begin{IEEEkeywords}
		Cell Free Massive MIMO, Deep Learning, Channel Reciprocity
	\end{IEEEkeywords}

	\section{Introduction}
Cell Free MIMO is a potential paradigm shift in wireless network design for 5G and 
beyond that includes the benefits of Massive MIMO as well as ability to exploit diversity 
and increase the immunity against shadow fading \cite{marzetta2017cellfree}. However, 
the technology is still limited by many practical constraints such as channel 
non-reciprocity \cite{marzetta2017cellfree}.

Even though most MIMO systems operate in TDD mode, the end to end channel is typically 
rendered \textit{non-reciprocal} because of the RF front ends and usually careful 
calibration is required to achieve reciprocity \cite{vieira2017reciprocity}. 
The impact of non-reciprocal channels on the gains of Cell Free Massive MIMO 
is analysed in \cite{impact}.
Even in the presence of reciprocity, the acquisition of the complete Downlink CSI 
from the Uplink CSI is not very straightforward, due to the non availability of Uplink 
CSI in the subcarriers which don't have reference signals.

Popular approaches for TDD reciprocity calibration are based on internal 
BS sounding using dedicated RF circuitry or over the air sounding among the Access 
Points(APs) which requires coordination \cite{vieira2017reciprocity,shepard2012argos,
jiang2018framework}. Most of these methods are only relative calibration methods 
that achieve reciprocity up to a multiplicative constant, which needs to be estimated 
\cite{shepard2012argos}. The main limitation on application of traditional calibration 
methods to a Cell Free Massive MIMO setting is the need for over the air sounding 
and requirement of stringent synchronization between the APs, which often is very 
hard since there is no centralized reference clocks for these APs. Hence, it is desirable 
to have an approach in which each AP has full control over it's local CSI, rather 
than relying on the other APs to acquire its CSI.

Most works in Cell Free Massive MIMO also focus on narrowband communication and assume 
that pilots are only multiplexed in the code domain as orthogonal reference signals. 
But, many cellular standards such as 5G, adopt Multicarrier schemes and use pilot 
aided methods for channel estimation, where the pilots are multiplexed in both the 
code domain as well as the  frequency domain \cite{dahlman20185g}. In a comb type 
pilot structure, reference signals (pilots) are inserted at specific subcarriers 
in the grid, and the pilot signals are frequency multiplexed among users 
\cite{dahlman20185g}. It is possible to estimate the CSI at the pilot subcarriers 
using standard estimation methods. However, at the subcarriers that do not have reference 
signals, which we call as \textit{blind subcarriers}, it is difficult to estimate 
highly accurate CSI without any prior knowledge, such as the second order statistics 
for an MMSE estimator \cite{coleri2002channel}. Hence, limited resource elements 
in reference signals for channel estimation makes it hard to obtain accurate CSI 
required for achieving high datarates.

A popular approach to combat the non-availability of CSI at blind subcarriers is 
frequency domain interpolation. But linear interpolation methods typically require 
dense pilots, which in turn reduces the spectral efficiency. Thus it might be desirable 
to have powerful non-linear interpolators. The combination of Reciprocity calibration 
as well as Frequency Domain interpolation is clearly a nonlinear problem, when assuming 
that pilots are sparse.

Artificial Neural Networks (ANNs) are widely used as non-linear function approximators 
and application of deep learning methods to problems in wireless communication systems 
has become popular recently due to its competitive performance. Deep learning has 
been successfully applied in the problems of communication systems design 
\cite{o2017introduction,ye2018power,raj2018backpropagating}, OFDM systems 
\cite{gao2018comnet} etc. In this work, we present a cascaded deep learning based 
method for inter-pilot interpolation and TDD reciprocity calibration in Cell Free 
Massive MIMO systems.

Major contributions of this work are
\begin{enumerate}
    \item A novel combined method for TDD-reciprocity calibration and CSI interpolation 
    using deep learning to recover the Downlink CSI across the entire Bandwidth part 
    from the Uplink CSI obtained at a small number of pilot subcarriers.
    \item A scalable and intelligent system for Cell Free Massive MIMO that identifies 
    the frequency selectivity of the environment of operation and performs accurate 
    interpolation accordingly.
\end{enumerate}

\subsection{Notations}
    Bold face upper case (eg. \mymat{A}) bold lower case letters denotes $\myvec{b}$
    matrix and column vectors respectively. Inverse of a matrix $\mymat{A}$ is denoted by
    $\mymat{A}^{-1}$ and transpose by $\mymat{A}^{T}$. Element at $i^{th}$ row $j^{th}$ column
    of matrix $\mymat{A}$ is denoted by $a^{(ij)}$.
	\section{System Model} \label{sec:sys_model}
Consider a multicarrier system of $N$ subcarriers with  $M$ APs and $K$ UEs. There 
are a total of $M \times K$ channels for each of the $n$ ($n = 1,2,\ldots,N$) subcarriers. 
We assume perfect synchronization and coordinated communication.

Let $\bf{x}_{UL} \in \mathbb{C}^{K \times 1}$ and $\bf{x}_{DL}  \in \mathbb{C}^{M 
\times 1}$ be the transmit symbols on the one subcarrier during uplink and downlink 
respectively. The received symbols for uplink and downlink can then be written as
\begin{align}
    \myvec{y}_{UL} &= \mymat{H}_{UL} \myvec{x}_{UL} + \myvec{w}_{UL} \\ 
    \myvec{y}_{DL} &= \mymat{H}_{DL} \myvec{x}_{DL} + \myvec{w}_{DL},
\end{align}
where $\bf{H}_{UL} \in \mathbb{C}^{M \times K}$ and $\bf{H}_{DL} \in \mathbb{C}^{K 
\times M}$ are the uplink and downlink channels respectively for that subcarrier 
and $\bf{w}$ is the corresponding AWGN noise. The channels $\bf{H}_{UL}$ and $\bf{H}_{DL}$ 
will include the effects from the RF front-end at both the transmitter and receiver. 
These effects can be captured using the model\cite{vieira2017reciprocity}
\begin{align}
 \mymat{H}_{UL} &= \mymat{R}_{UL} \mymat{C}_{UL} \mymat{T}_{UL}  \label{eqn:h_uplink}\\
 \mymat{H}_{DL} &= \mymat{R}_{DL} \mymat{C}_{DL} \mymat{T}_{DL}  \label{eqn:h_downlink}
\end{align}
where $\mymat{T}_{UL} \in \mathbb{C}^{K \times K}$, $\mymat{R}_{DL} \in \mathbb{C}^{K 
\times K}$, $\mymat{T}_{DL} \in \mathbb{C}^{M \times M}$ and $\mymat{R}_{UL} \in 
\mathbb{C}^{M \times M}$ and are the RF transmitter and receiver chains of the UE 
and BS respectively. The diagonal elements in $\mymat{T}_{DL},\mymat{R}_{DL},
\mymat{T}_{UL},\mymat{R}_{UL}$ correspond to the gains of individual chains and the 
off-diagonal elements corresponds to RF-cross talk and antenna coupling.  
Here, $\mymat{C}_{UL}$ and $\mymat{C}_{DL}$ are the physical wireless channels between 
the RF front-ends of BS and UE. The $(m,k)^{th}$ element of $\mymat{C}_{UL}$, $g_{m,k}$ 
is modelled as \cite{marzetta2017cellfree}, $g_{m,k} = \sqrt{\beta_{m,k}}h_{m,k}$, 
where $\beta_{m,k}$ represents the large scale fading and $h_{m,k}\sim\mathcal{CN}(0,1)$
represents the small scale fading.

We assume that the RF chains are Linear Time Invariant (LTI).
Assuming TDD mode of operation, the wireless channel is reciprocal for every link 
between BS and UE. Hence, we have
$
    \mymat{C}_{DL} = (\mymat{C}_{UL})^T.
$
Under the assumption the RF front end matrices are invertible, the downlink channel
$\mymat{H}_{DL}$ can be decomposed as
\begin{align}
    \mymat{H}_{DL} 
        &= \mymat{R}_{DL} \mymat{C}_{DL} \mymat{T}_{DL} \nonumber \\
        &= \mymat{R}_{DL} (\mymat{R}_{UL}^{-1} \mymat{H}_{UL} 
                \mymat{T}_{UL}^{-1})^T \mymat{T}_{DL} \nonumber \\
        &= \mymat{R}_{DL} (\mymat{T}_{UL}^{-1})^T (\mymat{H}_{UL})^T
                (\mymat{R}_{UL}^{-1})^T \mymat{T}_{DL}
                \label{eqn:recip_mimo}
\end{align}
Hence, with channel reciprocity, the downlink channel $\mymat{H}_{DL}$ can be
computed as a transformation of the uplink channel $\mymat{H}_{UL}$. However, for the estimation
of $\mymat{H}_{DL}$, perfect knowledge of RF front end matrices are required. In practical
cases,  this information is not easily available and traditional methods resort to internal 
sounding based techniques \cite{vieira2017reciprocity,shepard2012argos,jiang2018framework}.

\subsection{OFDM based Cell Free Massive MIMO}
Orthogonal Frequency Division Multiplexing (OFDM) is a popular multicarrier scheme 
that is used in current wireless standards such as 5G-NR and 4G-LTE. We consider 
a Cell Free Massive MIMO based system with OFDM scheme. At an individual single antenna 
transmitter, an OFDM frame is built by inserting pilots in to the data-block and 
then taking inverse discrete Fourier transform (IDFT) to convert the signal from 
frequency domain to time domain. Then a cyclic prefix (CP), of length no shorter 
than delay spread, is inserted before transmission. At a single AP $m$, the transmitted 
frequency domain signal in the Downlink to a single UE $k$ can be represented as
\begin{align}
    y_{DL}(n) = g_{mk,DL}(n) x_{DL}(n) + w_{DL}(n), \quad n = 1, \ldots, N,
\end{align}
where $N$ is the block length (number of sub-carriers) of OFDM block. The precoding 
for a Downlink OFDM block requires $g_{mk,DL}$ at each sub-carrier, which needs to 
be obtained from the Uplink CSI. The absence of perfect CSI due to channel estimation 
error or non-reciprocity degrades the performance of Massive MIMO and could inhibit 
one from realizing it's full potential \cite{mi2017massive}. Assuming we know the 
Uplink CSI at the pilot subcarriers, the Downlink CSI at all the blind subcarriers 
need to be estimated.

For a single link between the $m^{th}$ AP and the $k^{th}$ UE, the Downlink channel 
$g_{mk,DL}$ at the $nth$ subcarrier could be written as:
\begin{align}
    g_{mk,DL}(n) = f\left(\frac{r_{DL}(l)t_{DL}(l)}{r_{UL}(l)t_{UL}(l)}g_{mk,UL(l)}\right) \label{eq:single}
\end{align}
Here the function $f(.)$ between channel at a blind subcarrier $n$ and a pilot subcarrier 
$l$ is induced mainly by the wireless channel, which is nothing but a measure of frequency 
selectivity of the channel.
    \section{Proposed Approach}
\revAdd{From (\ref{eq:single}), we can see that for a single channel between a UE 
and an AP, process of predicting the Downlink CSI at all sub-carriers from the Uplink 
channel estimates at pilot positions involves both reciprocity calibration and frequency 
domain interpolation. Using Deep learning, one could approximate the downlink channel 
at all subcarriers from the uplink channel at pilot subcarriers. Even though the 
RF chain responses could be assumed to be roughly constant for a long time \cite{impact}, 
the function induced by the wireless channel depends on the scenario. Modern wireless 
standards require the devices to operate in multiple scenarios of operation such 
as Indoor Hotspot, Urban, Rural, etc and the frequency selectivity differs across 
scenarios. Since it is essential that an interpolation function works for every scenario, 
one needs to learn the type of frequency selectivity in the current scenario. We 
assume that there are $S$ classes of channels, with each class having different Power 
Delay Profile (PDP).

Hence, (\ref{eq:single}) now becomes:
\begin{align}
    g_{mk,DL}(n) = f_s\left(\frac{r_{DL}(l)t_DL(l)}{r_{UL}(l)t_UL(l)}g_{mk,UL(l)}\right)\bm{i}(g_{mk,UL}\in s)\label{eq:ind}
\end{align}
where \begin{math}\bm{i}(g_{mk,UL})\end{math}  $\in\{0,1\}$ is an indicator variable 
to indicate which of the $S$ classes the channel belongs to. The indicator variable 
for the $sth$ class is 1 and the variables for the other $S-1$ classes are 0 if the 
channel $\bm{i}(g_{mk,UL})$  belongs to the $s^{th}$ class.}

\subsection{Channel Identification}
Since the frequency domain correlation of the channel is the fourier transform of 
the PDP, it can be understood that PDP determines the function $f_s(.)$ in (\ref{eq:ind}). 
Since the PDP changes across scenarios, the first step in the  proposed method is 
to classify the Uplink channel estimates into one of the known channel classes/scenarios 
based on the PDP using the CSI from pilot positions. The classifier performs the 
role of the indicator.

We assume that each block of the Uplink transmitted data is inserted with $N$ pilots. 
Upon reception of the block, we can obtain the CSI at each of these pilot positions 
through any of the standard channel estimation methods \cite{coleri2002channel}. 
This information is then fed to a neural network based classifier for channel 
identification. Since neural networks can only work with real numbers, we flatten 
the complex CSI information into real and imaginary parts and feed it to the classifier 
network. Thus the neural network takes in an input vector of dimension $2N$. The 
output layer of the network is of dimension $S$ with softmax activation function, 
computing a surrogate probability of the provided sample being in each of the class. 
During training phase, samples from different channel scenarios are used with corresponding 
one-hot labeling for class. Categorical cross entropy is used as the loss function 
to train the classifier.

\subsection{Interpolation and Reciprocity Calibration}
We combine the process of interpolation and TDD reciprocity calibration into one 
step and train a deep neural network (DNN) with this objective. We train $S$ different 
DNNs for this purpose; one for each of $S$ class of channels identified during the 
system modelling. Each DNN is trained to take a $2N$ dimensional input, corresponding 
to the flattened CSI information from the pilot positions. The output is of dimension 
$2K$ corresponding to the real and imaginary parts of the downlink channel which 
includes both the calibration and interpolation. During training, the CSI information 
along with the actual downlink channel information from the corresponding channel 
class is fed to the network. We used mean squared error as the loss function to train 
the network.

\begin{figure}[!h]
    \centering
    \tikzset{dotted_block/.style={draw=black!80!white, line width=1pt, 
            dash pattern=on 1pt off 4pt on 6pt off 4pt,
            inner sep=8mm, 
            rectangle, 
            rounded corners}}
            
\begin{tikzpicture}[auto,>=stealth',every text node part/.style={align=center},scale=0.70, every node/.style={transform shape}]
    \begin{scope}[auto,node distance=1.0cm]
        \node [draw,fill=white,rectangle,minimum height=0.8cm,minimum width=2.0cm] at (0.0,0.0)
            (rx_block) {Uplink Received Block};
            
        \node [draw,fill=white,rectangle,minimum height=0.8cm,minimum width=2.0cm] at (4.0,0.0)
            (Channel Estimation) {Channel Estimation};
        \draw[->] (rx_block) -- (Channel Estimation);
            
        \node [draw,fill=white,rectangle,minimum height=0.8cm,minimum width=2.0cm] at (8.0,0.0)
            (classifier_dnn) {Channel Classifier};
        \draw[->] (Channel Estimation) -- (classifier_dnn);
            
        \node [draw,fill=white,rectangle,minimum height=0.8cm,minimum width=2.0cm] at (8.0,-2.0)
            (calib_dnn_1) {Calibration DNN 1};
        \node [draw,fill=white,rectangle,minimum height=0.8cm,minimum width=2.0cm] at (8.0,-3.0)
            (calib_dnn_2) {Calibration DNN 2};
        \draw [dotted] (calib_dnn_2) -- (8.0,-4.4);
        \node [draw,fill=white,rectangle,minimum height=0.8cm,minimum width=2.0cm] at (8.0,-5.0)
            (calib_dnn_s) {Calibration DNN $S$};
        \node (calib_block) [dotted_block, fit = (calib_dnn_1) (calib_dnn_1) (calib_dnn_s)] {};
        \draw[->] (classifier_dnn) -- (calib_block);
        
        \node [draw,fill=white,rectangle,minimum height=0.8cm,minimum width=2.0cm] at (2.0,-3.5)
            (dl_predicted) {Predicted DL channel};
        \draw[->] (calib_block) -- (dl_predicted);
    \end{scope}
\end{tikzpicture}
    \caption{Schematic of the proposed TDDNet scheme.}
    \label{fig:proposed_schematic}
\end{figure}
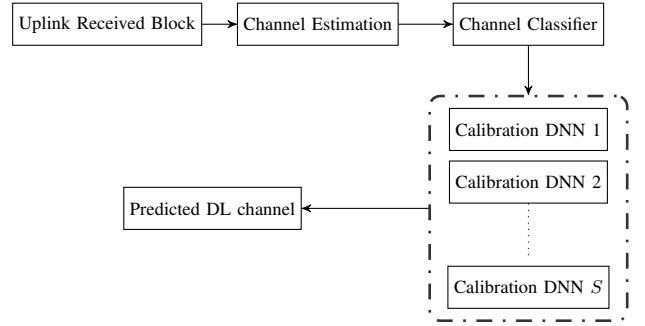
A schematic representation of the proposed \textit{TDDNet} approach is given in Fig. 
\ref{fig:proposed_schematic}. In a practical situation, the CSI information that 
can be estimated includes the wireless channel impairments and RF impairments at 
both transmitter and receiver. Hence, we propose to use the data which includes these 
impairments as this the closest we can get to practical scenarios. More specifically, 
we use the UL CSI which includes the RF impairments at both the AP and UE to predict 
the DL CSI which also includes the RF impairments at both the sides as the input.
During the training phase, the classifier network and the calibration DNNs are individually 
trained. During testing, the samples are first fed to classifier network, which activates 
one of the calibration networks based on the detected channel profile and the downlink 
CSI is predicted for all subcarriers by the model. An algorithmic description of 
the proposed method is provided in Alg. \ref{alg:proposed}. 

\begin{algorithm}[!t]
    \caption{Algorithm for Downlink Channel prediction}
    \label{alg:proposed}
    \begin{algorithmic}[1]
        \State \textbf{Input:} Received uplink pilot block $y_{P,UL}$
        \State \textbf{Output:} Downlink CSI $h_{DL}$ 
        \For {Each received block}
            \State Compute Uplink channel at pilot positions as $h_{P,UL}(n) = y_{UL}(n) / x_{UL}(n)$ where $x_{UL}(n)$ is the known pilot symbol  at the $nth$ subcarrier and $h_{P,UL}$ is a vector of uplink channel estimates at pilot positions.
            \State Classify $h_{P,UL}$ to one of the known channel classes using the classifier network. \revDel{It should be noted that the power delay profile remains the same in the uplink and the downlink since the physical wireless channel remains the same.}
            \State Select appropriate downlink channel prediction DNN based on the output of the classifier.
            \State Use $h_{P,UL}$ and the selected DNN to predict the downlink channel $h_{DL}$ for all subcarriers.
        \EndFor
    \end{algorithmic}
\end{algorithm}

\subsection{Scaling up for Multiple links}
In a Cell Free scenario, there are multiple links which can be assumed to be i.i.d 
due to the distribution of APs and UEs. In this case, our method is easily scalable 
since each AP does not have to depend on the other APs for calibration, in contrary 
to the traditional relative calibration schemes in which each AP has to coordinate 
with other APs to achieve reciprocity. Usually traditional calibration is achieved 
by having a reference RF chain among the operating RF chains and sounding calibration 
signals to the reference chain \cite{marzetta2010noncooperative}\cite{shepard2012argos}. 
By obviating the dependence of APs, over the air transmission of reference signals 
among APs is not required. This means that as and when new APs are added to the network, 
there is no disruption caused to it by the operating APs. Such an independence is 
crucial to the flexibility offered by Cell Free systems.

The proposed method of calibration on a per link basis also applies to multiple UEs 
since each UE might have a unique RF chain response and each AP is required to calibrate 
reciprocity individually with each of the UEs.

    \section{Experimental Results}
This section presents the results comparing the proposed method with popular approaches 
present in literature. For evaluation, we have identified 5 channel classes based 
on (3GPP) TR 38.901 Release 15 \cite{3gpp_ts_38901_5g} Channel models viz.: TDL-A, 
TDL-B, TDL-C, TDL-D, TDL-E. We followed Sec 7.7.5.2 of TR 38.901 Release 15 to use 
the TDL models for MIMO channels.  Each of these classes has different delay profile 
and thus a different frequency structure. We consider $256$ subcarriers with $30$KHz 
subcarrier spacing. The carrier frequency is set to $3.5$GHz and the sampling frequency 
is $100$MHz. For modelling a moderate time selectivity, we used a UE velocity of 
$20$kmph for all simulations. The gains of RF chains across subcarriers has been 
modelled as i.i.d random variables distributed as $X \sim \mathcal{CN}(g_m,\sigma^{2})$ 
(where $g_m$ is the average baseband gain of the RF chain and $\sigma^{2}$ is the variance), 
and are kept constant throughout the experiment. The wireless channels and RF chains 
were generated using MATLAB and the 5G toolbox of MATLAB.

During the training phase, samples of the Uplink CSI from pilot positions from different 
channel models are labeled and used to train the classifier network. For comparing 
the MSE performance of the proposed approach across SNR, we used a system with $256$ 
subcarriers out of which $11$ are pilots. We split the complex Uplink CSI information 
into two real numbers and stack them together to form a $22$ dimensional input vector 
for the system. Similarly, the output from the network is a $512$ dimensional real 
vector which we reshape into a $256$ dimensional complex vector for obtaining the 
downlink CSI. The details for training classifier network is given in Table 
\ref{tab:classifier_param}. The Downlink channel prediction network is trained specifically, 
one for each channel class. Details for training the network are given in  Table 
\ref{tab:recipro_param}.

\begin{table}[!h]
    \centering
    \begin{tabular}{|l|c|}
        \hline 
        \multicolumn{1}{|c|}{\textbf{Parameter}} & \multicolumn{1}{|c|}{\textbf{Value}} \\
        \hline
        \hline
        Input dimension  & $22$           \\
        Hidden Layer 1   & $22$ (tanh)    \\
        Hidden Layer 2   & $22$ (sigmoid) \\
        Output dimension & $5$ (softmax)  \\
        Loss             & Cross Entropy  \\
        \hline
    \end{tabular}
    \caption{Parameters for classifier network} \label{tab:classifier_param}
\end{table}

\begin{table}[!h]
    \centering
    \begin{tabular}{|l|c|}
    \hline
    \multicolumn{1}{|c|}{\textbf{Parameter}} & \multicolumn{1}{|c|}{\textbf{Value}} \\
    \hline
    \hline
    Input Dimension                        & 22 \\
    Hidden Layer 1                         & $512$ (tanh) \\
    Hidden Layer 2                         & $128$ (tanh) \\
    Output Layer                           & $512$ (linear)  \\ 
    Loss                                   & MSE \\
    \hline
    \end{tabular}
    \caption{Parameters for channel prediction network} \label{tab:recipro_param}
\end{table}

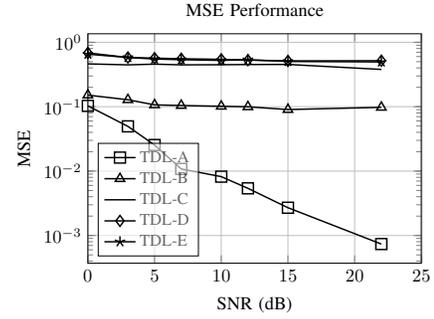
\begin{figure}[!h]
    \centering
    \resizebox{0.65\linewidth}{!}{\begin{tikzpicture}[thick,scale=0.8]
    \begin{semilogyaxis}[
        width=8cm,
        height=6cm,
        grid=major,
        xmin=0,
        xmax=25,
        xlabel={SNR (dB)},
        ylabel={MSE},
        xlabel style={at={(0.50,0.00)}},
        ylabel style={at={(0.00,0.50)}},
        title={MSE Performance},
        legend pos=south west,
        legend cell align={left},
        legend style={fill opacity=0.6, draw opacity=1.0, text opacity=1.0, minimum width=1cm, font=\small}
        ]
        
        \addplot[black, solid, thick, mark=square, mark size={3.0}] 
            table [x=True_TDL_A_x, y=True_TDL_A_y, col sep=comma]{./data/mse_mismatch.csv};
        \addlegendentry{TDL-A};
        
        \addplot[black, solid, thick, mark=triangle, mark size={3.0}] 
            table [x=TDL_B_x, y=TDL_B_y, col sep=comma]{./data/mse_mismatch.csv};
        \addlegendentry{TDL-B};
        
        \addplot[black, solid, thick, mark=circle, mark size={3.0}] 
            table [x=TDL_C_x, y=TDL_C_y, col sep=comma]{./data/mse_mismatch.csv};
        \addlegendentry{TDL-C};
        
        \addplot[black, solid, thick, mark=diamond, mark size={3.0}] 
            table [x=TDL_D_x, y=TDL_D_y, col sep=comma]{./data/mse_mismatch.csv};
        \addlegendentry{TDL-D};
        
        \addplot[black, solid, thick, mark=star, mark size={3.0}] 
            table [x=TDL_E_x, y=TDL_E_y, col sep=comma]{./data/mse_mismatch.csv};
        \addlegendentry{TDL-E};
        
    \end{semilogyaxis}
\end{tikzpicture}}
    \caption{MSE when all channels are predicted using TDL-A model.}
    \label{fig:mismatch_curve}
\end{figure}

The first step in the proposed approach is the classification of observed channel 
samples into one of the pre-identified classes. Fig. \ref{fig:mismatch_curve} highlights 
the necessity of an accurate classifier at the first stage, and justifies the cascade 
approach. In this experiment, a DNN trained for TDL-A channel model is used for predicting 
the downlink channel for different channel models. Without a classifier, the mismatch 
in the PDP increases the MSE as shown in the figure.  Similar observations can also 
be made for other channel models also.

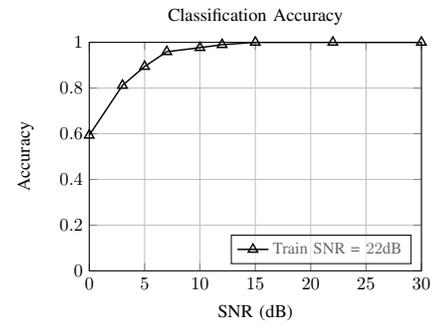
\begin{figure}[!h]
    \centering
    \resizebox{0.65\linewidth}{!}{\begin{tikzpicture}[thick,scale=0.8]
    \begin{axis}[
        width=8cm,
        height=6cm,
        xmin=0,
        xmax=30,
        ymin=0.00,
        ymax=1.00,
        grid=major,
        xlabel={SNR (dB)},
        ylabel={Accuracy},
        xlabel style={at={(0.50,0.00)}},
        ylabel style={at={(0.00,0.50)}},
        ytick={0.00,0.20,...,1.00},
        log ticks with fixed point,
        title={Classification Accuracy},
        legend pos=south east,
        legend cell align={left},
        legend style={fill opacity=0.6, draw opacity=1.0, text opacity=1.0, minimum width=1cm, font=\small}
        ]
        
        \addplot[black, solid, thick, mark=triangle, mark size={3.0}] 
            table [x=data1_x, y=data1_y, col sep=comma]{./data/snr_acc.csv};
        \addlegendentry{Train SNR = $22$dB};
        
    \end{axis}
\end{tikzpicture}}
    \caption{Accuracy plot classifer network at different SNR (dB).}
    \label{fig:fig_acc}
\end{figure}

As the first part of the cascaded architecture, the classifier network needs to have 
high accuracy to correctly classify the channel samples across a wide range of SNR. 
Fig. \ref{fig:fig_acc} provides the classification accuracy of the trained network 
across different SNR. At low SNR, the signal strength received may not be enough 
to correctly identify the Power Delay Profile (PDP) of the channel and hence we see 
a drop in accuracy at lower SNRs. However, even in low SNR, the trained network is 
able to correctly classify the PDPs of the channel with more that $60\%$ accuracy.

\begin{figure*}[!t]
    \centering
    \begin{subfigure}{.33\textwidth}
        \resizebox{\linewidth}{!}{\begin{tikzpicture}[thick,scale=0.8]
    \begin{semilogyaxis}[
        width=8cm,
        height=6cm,
        grid=major,
        xmin=0,
        xmax=25,
        xlabel={SNR (dB)},
        ylabel={MSE},
        xlabel style={at={(0.50,0.00)}},
        ylabel style={at={(0.00,0.50)}},
        title={MSE Performance},
        legend pos=north east,
        legend cell align={left},
        legend style={fill opacity=0.6, draw opacity=1.0, text opacity=1.0, minimum width=1cm, font=\small}
        ]
        
        \addplot[black, solid, thick, mark=square, mark size={3.0}] 
            table [x=TDL_A_x, y=TDL_A_y, col sep=comma]{./data/mse_snr_cascade.csv};
        \addlegendentry{TDL-A};
        \addplot[black, dotted, thick, mark=square, mark size={3.0}, mark options=solid, forget plot] 
            table [x=TDL_A_x, y=TDL_A_y, col sep=comma]{./data/mse_snr_oracle.csv};
        
        \addplot[black, solid, thick, mark=triangle, mark size={3.0}] 
            table [x=TDL_B_x, y=TDL_B_y, col sep=comma]{./data/mse_snr_cascade.csv};
        \addlegendentry{TDL-B};
        \addplot[black, dotted, thick, mark=triangle, mark size={3.0}, mark options=solid, forget plot] 
            table [x=TDL_B_x, y=TDL_B_y, col sep=comma]{./data/mse_snr_oracle.csv};
        
        \addplot[black, solid, thick, mark=circle, mark size={3.0}] 
            table [x=TDL_C_x, y=TDL_C_y, col sep=comma]{./data/mse_snr_cascade.csv};
        \addlegendentry{TDL-C};
        \addplot[black, dotted, thick, mark=circle, mark size={3.0}, mark options=solid, forget plot] 
            table [x=TDL_C_x, y=TDL_C_y, col sep=comma]{./data/mse_snr_oracle.csv};
        
        \addplot[black, solid, thick, mark=diamond, mark size={3.0}] 
            table [x=TDL_D_x, y=TDL_D_y, col sep=comma]{./data/mse_snr_cascade.csv};
        \addlegendentry{TDL-D};
        \addplot[black, dotted, thick, mark=diamond, mark size={3.0}, mark options=solid, forget plot] 
            table [x=TDL_D_x, y=TDL_D_y, col sep=comma]{./data/mse_snr_oracle.csv};
        
        \addplot[black, solid, thick, mark=star, mark size={3.0}] 
            table [x=TDL_E_x, y=TDL_E_y, col sep=comma]{./data/mse_snr_cascade.csv};
        \addlegendentry{TDL-E};
        \addplot[black, dotted, thick, mark=star, mark size={3.0}, mark options=solid, forget plot] 
            table [x=TDL_E_x, y=TDL_E_y, col sep=comma]{./data/mse_snr_oracle.csv};
        
    \end{semilogyaxis}
\end{tikzpicture}}
        \caption{MSE in prediction}
        \label{fig:fig_mse}
    \end{subfigure}%
    \begin{subfigure}{.33\textwidth}
        \resizebox{\linewidth}{!}{\begin{tikzpicture}[thick,scale=0.8]
    \begin{semilogyaxis}[
        width=8cm,
        height=6cm,
        grid=major,
        xmin=0,
        xmax=40,
        xlabel={Pilot Spacing (subcarriers)},
        ylabel={MSE},
        xlabel style={at={(0.50,0.00)}},
        ylabel style={at={(0.00,0.50)}},
        title={Effect of Pilots},
        legend pos=north west,
        legend cell align={left},
        legend style={fill opacity=0.6, draw opacity=1.0, text opacity=1.0, minimum width=1cm, font=\small}
        ]
        
        \addplot[black, solid, thick, mark=triangle, mark size={3.0}] 
            table [x=DNN_x, y=DNN_y, col sep=comma]{./data/mse_pilot.csv};
        \addlegendentry{Proposed};
        
        \addplot[black, solid, thick, mark=square, mark size={3.0}] 
            table [x=Wiener_filter_x, y=Wiener_filter_y, col sep=comma]{./data/mse_pilot.csv};
        \addlegendentry{Wiener Filter};
        
        \addplot[black, solid, thick, mark=star, mark size={3.0}] 
            table [x=Linear_x, y=Linear_y, col sep=comma]{./data/mse_pilot.csv};
        \addlegendentry{Linear};
        
    \end{semilogyaxis}
\end{tikzpicture}}
        \caption{Effect of pilot spacing}
        \label{fig:fig_pilots}
    \end{subfigure}%
    \begin{subfigure}{.33\textwidth}
        \resizebox{\linewidth}{!}{
\def\mystrut{\vphantom{hg}}

\pgfplotsset{
    legend image with text/.style={
        legend image code/.code={%
            \node[anchor=center] at (0.3cm,0cm) {#1};
        }
    },
}
\begin{tikzpicture}[thick,scale=0.8]
    \begin{semilogyaxis}[
        name=snrplot,
        width=8cm,
        height=6cm,
        grid=major,
        xmin=0,
        xmax=25,
        xlabel={SNR (dB)},
        ylabel={MSE},
        xlabel style={at={(0.50,0.00)}},
        ylabel style={at={(0.00,0.50)}},
        title={MSE v/s SNR},
        legend pos=north east,
        legend cell align={left},
        legend columns=2,
        legend style={fill opacity=0.6, draw opacity=1.0, text opacity=1.0, font=\small}
        ]
        
        \addlegendimage{legend image with text=TDL-A};
        \addlegendentry{};
        
        \addlegendimage{legend image with text=TDL-E};
        \addlegendentry{};

        \addplot[black, solid, thick, mark=triangle, mark size={3.0}] 
            table [x=TDLA_DNN_x, y=TDLA_DNN_y, col sep=comma]{./data/mse_snr_compare.csv};
        \addlegendentry{};
        \label{plot:tdla_proposed};
        
        \addplot[black, dashed, thick, mark=triangle, mark size={3.0}, mark options=solid] 
            table [x=TDLE_DNN_x, y=TDLE_DNN_y, col sep=comma]{./data/mse_snr_compare.csv};
        \addlegendentry{Proposed};
        \label{plot:tdle_proposed};
        
        \addplot[black, solid, thick, mark=square, mark size={3.0}] 
            table [x=TDLA_Wiener_x, y=TDLA_Wiener_y, col sep=comma]{./data/mse_snr_compare.csv};
        \addlegendentry{};
        \label{plot:tdla_wiener};
        
        \addplot[black, dashed, thick, mark=square, mark size={3.0}, mark options=solid] 
            table [x=TDLE_Wiener_x, y=TDLE_Wiener_y, col sep=comma]{./data/mse_snr_compare.csv};
        \addlegendentry{Wiener};
        \label{plot:tdle_wiener};
        
        \addplot[black, solid, thick, mark=star, mark size={3.0}] 
            table [x=TDLA_Linear_x, y=TDLA_Linear_y, col sep=comma]{./data/mse_snr_compare.csv};
        \addlegendentry{};
        \label{plot:tdla_linear};
        
        \addplot[black, dashed, thick, mark=star, mark size={3.0}, mark options=solid] 
            table [x=TDLE_Linear_x, y=TDLE_Linear_y, col sep=comma]{./data/mse_snr_compare.csv};
        \addlegendentry{Linear};
        \label{plot:tdle_linear};
    \end{semilogyaxis}
    
\end{tikzpicture}}
        \caption{Effect of SNR}
        \label{fig:fig_snr}
    \end{subfigure}%
    \caption{Performance of proposed approach}
    \label{fig:comparison_curve}
\end{figure*}
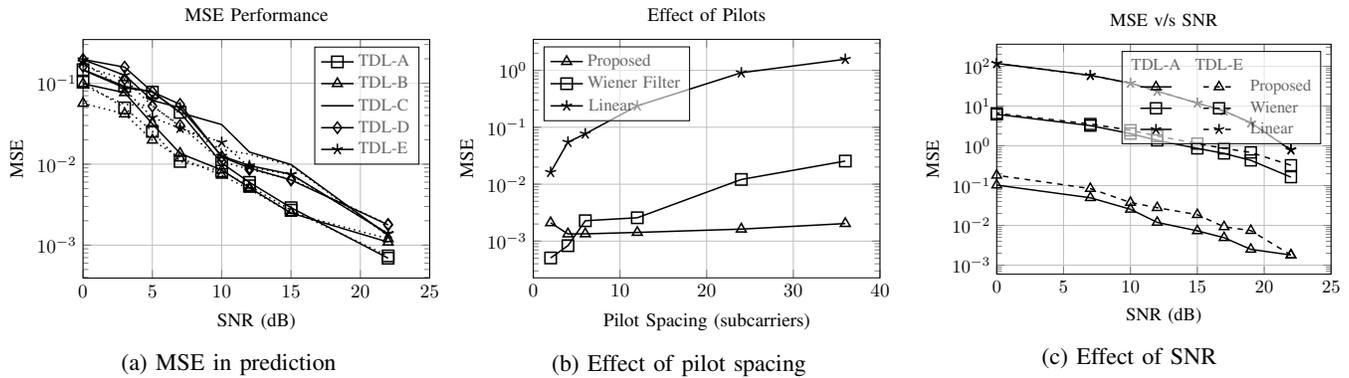

A comparison of results of the proposed system is given in Fig. \ref{fig:comparison_curve}. 
The MSE of the proposed cascaded approach for different channel models are given 
in Fig. \ref{fig:fig_mse}. A pilot spacing of 24 subcarriers is used. We can observe 
the proposed method shows similar trend in performance for all channel models. The 
dotted line is the MSE performance of an oracle classifier which always classifies 
the channel samples correctly and activates the \textit{correct} interpolation/calibration 
DNN for Downlink channel prediction. We can observe that the MSE performance of oracle 
is slightly better than the proposed method at lower SNRs. The accuracy of classifier 
is not $100\%$ at lower SNRs and this why the proposed cascade method incurs slightly
higher MSE at that regime.

The effect of Uplink  pilot spacing for Downlink channel prediction is studied in 
Fig. \ref{fig:fig_pilots} for the TDL-C channel type at 22 dB SNR. The proposed method 
is compared against popular Linear interpolation method  and Wiener Filter (Linear 
MMSE assuming perfect knowledge about second order statistics of the channel) based 
interpolation method. We can observe that both Linear and Wiener Filter based interpolation 
are sensitive to Uplink pilot spacing while the proposed method is robust to this.

The MSE performance of the methods under comparison with sparse pilot scenario is 
given in Fig. \ref{fig:fig_snr}. We used a pilot spacing of $24$ in $256$ subcarrier 
system. This results in a total of $11$ pilots. It can be clearly observed that the 
proposed method provides better estimation of downlink channel even with sparse pilots, 
while the traditional methods (Linear and Wiener filter methods) suffer with high MSE. 
This improvement in MSE even in sparse pilot scenario can be attributed to the capabilities 
of neural networks to perform high resolution approximation of non-linear functions.
    \section{Concluding Remarks}
We discussed the problem of Downlink CSI acquisition from the Uplink CSI in a Cell 
Free Massive MIMO scenario and proposed a method using Deep learning to solve it. 
We presented a method for reciprocity calibration and obtaining the complete channel 
estimate for precoding purposes using a cascade of DNNs. Results indicate that our 
method outperforms traditional methods significantly even with a few number of pilots, 
thus providing better spectral efficiency. Even though we discussed the utility of 
cascaded deep learning based channel estimation in the context of TDD Cell Free Massive 
MIMO, the proposed method can be applied to systems which otherwise require the User 
Equipment(UE)  to estimate downlink CSI.

\section{Acknowledgement}
The authors would like to thank Dr. Radhakrishna Ganti, Associate Professor of Electrical 
Engineering, IIT Madras for useful discussion.

	
 	\bibliographystyle{IEEEtran}
    \bibliography{library.bib}
\end{document}